# A LABORATORY STUDY OF ABSORBING CAPACITY OF WATER VAPOR AT THE WAVELENGTHS FROM 6500 TO 10500 Å

G.A.Alekseeva, V.D.Galkin, and I.B.Sal'nikov

The Central (Pulkovo) Astronomical Observatory, Russian Academy of Sciences, Russia

*We obtained laboratory spectra of absorption by water vapor at the wavelengths 6500-10500 Å with the multipass cell. The water vapor content along the line of view varied from 0.1 to 3.0 cm of precipitated water, the pressure from 0.1 to 1.0 atm. The spectra were taken with the width of the exit slit of the spectrophotometer 25, 50, 100, and 150 Å. To match these spectra, we selected empirical functions, which approximate the observed absorption within the indicated interval of water vapor content and pressure with the accuracy about 1%. For the water vapor band at the wavelengths regions 7200, 8200, and 9300 Å, with the step 25 Å, we determined the parameters necessary for the calculation of empirical transmission functions. The presented data make it possible to select the parameters for taking into account the radiation attenuation in the spectral region of telluric water vapor under the conditions of real astronomical observations for a specific place and spectrophotometer. The suggested set of empirical parameters may provide correction of observed stellar spectra for the extinction in the atmosphere with the accuracy $0.^m01-0.^m02$.*

In ground-based spectrophotometrical observations of stars, it is particularly important to properly reduce the observational to extra-atmospheric stellar magnitudes, i.e. to introduce corrections for the atmospheric extinction, which may cause serious errors in the final results. To this end, a commonly accepted technique of data processing is used, which necessarily includes the determination of extinction parameters. With the extension of the wavelength interval of observations to the IR region, the common problem of permanent testing of the stability of extinction was complicated by the supplementary problem of proper determination of the extinction in the portions of the spectrum that include bands of absorption by atmospheric water vapor. As an example of that, Fig.1 displays the transmission of ideal atmosphere $T_\lambda$ (in the absence of the aerosol component of the radiation extinction) at the wavelength interval 5500-10500 Å. The Figure shows that the alternation of the sequence of absorption bands makes the atmosphere to be a selective filter with complex spectral characteristics of transmission, which in turn vary in a complicated way, depending on the state of the atmosphere and its physical parameters.

In order to take into account correctly the impact of varying atmospheric transmission, the latter should be known for each moment of time of the observations for a given air mass. If the transmission of the atmosphere $T_\lambda$ for a moment of the observations is known, the correction for the atmospheric attenuation $\Delta m_\lambda$ is specified as

$$\Delta m_\lambda = -2.5 \lg T_\lambda, \qquad (1)$$

with the error of reduction

$$\delta(\Delta m_\lambda) = 1.086 \frac{\delta T_\lambda}{T_\lambda} \qquad (2)$$

It follows from the expression (2) that if the reduction accuracy is specified, then the higher is the transmission of the atmosphere, the lower are the requirements to the accuracy of its determination. And, inversely, if the transmission may be precisely determined, reliable results may be obtained even with low atmospheric transmission.

In spite of the fact that the absorption in the water vapor bands at the considered wavelengths was repeatedly studied both in laboratory and for the real atmosphere, the available data [1 - 6] indicate that currently, the correction of observations for the atmosphere does not make it possible to maintain the accuracy of 1-2%, which is commonly expected in photoelectrical observations. Here, we present our study of the spectral dependence of atmospheric transmission at the spectral region of water vapor bands, along with the experimentally obtained dependence of the transmission on the amount of absorbing matter and the pressure. We have also determined the accuracy with which the transmission may be approximated with the use of empirical functions. The study has been carried out on the basis of laboratory modeling of atmospheric absorption using the Pulkovo VKM-100 multipass cell. The experimental data were obtained for the values of physical parameters – the water vapor content, pressure, and temperature – close to the conditions of astronomical observations.

**The approximation of the transmission in the region of the absorption bands with the use of empirical parameter.**

In the considered spectral region, the water vapor absorption bands correspond to the wavelengths intervals 6900-7400, 8000-8400, 8900-10000 Å (Fig. 1). Within a band, the absorption is specified by the combined effect of the set of individual telluric lines, which fall within the wavelength interval discriminated by the slit of the spectrophotometer. The absorption in the lines depends on the amount of absorbing matter, pressure, and temperature. The transmission is also affected by the resolution of the instrument, i.e. the size of the wavelength interval $\Delta\lambda$ within which the transmission in individual lines is averaged. In the general case, the transmission may be presented by a function of the amount of the water vapor $\omega$ along the line of sight, the pressure P, the temperature T, and the spectral resolution $\Delta\lambda$:

$$T_\lambda = T_\lambda (\omega, P, T, \Delta\lambda) \qquad (3)$$

In order to calculate the transmission for a given set of physical conditions, a specific mathematical model for the dependence of the transmission on $\omega$, P, T, and $\Delta\lambda$ is needed. Different forms of such dependences were repeatedly discussed [7, 8]. Here, we used the expression

$$T_\lambda = \exp(-\beta_\lambda(T) \cdot \omega^{\mu_\lambda P^{n_\lambda}}), \qquad (4)$$

where $\beta_\lambda(T)$, $\mu_\lambda$, $n_\lambda$ are empirical parameters. The expression (4) was suggested in the study [9] as a model for the transmission of radiation in bands in a broad interval of the variation of the amount of absorbing matter along the line of sight, and was used in [4] for the reduction of spectrophotometrical observations of stars to extra-atmospheric magnitudes. Substituting the expression (4) into (1), for the attenuation in stellar magnitudes we obtain:

$$\Delta m_\lambda = C_\lambda \cdot \omega^{\mu}{}_\lambda, \qquad (5)$$

where $C_\lambda = -2.51 \lg e \cdot \beta_\lambda(T) \cdot P^n{}_\lambda$.

In the expression (5), the parameter $C_\lambda$ combines the parameters on which the transmission depends weakly and which may be considered constant for a given place of the observations. In the expression (5), the dependence of attenuation caused by the basic variable parameter, the amount of water vapor along the line of sight, is discriminated; in the course of observations, this parameter varies with both the variation of zenith distance of a star and that of the water vapor content in the atmosphere.

For a given place of the observations, parameter $C_\lambda$ corresponds to the transmission of the atmosphere for the water vapor content in the air column equal to 1 cm of precipitated water. This parameter indicates the rate of the variation of the attenuation depending on the amount of absorbing matter along the line of sight. The set of the parameters describes observed variation of the transmission at a given place of the observations with a given instrument, i.e. with a specific resolution $\Delta\lambda$. Here, we will neglect the temperature dependence of the transmission, which may result in the variations of the transmission in the considered wavelength interval that do not exceed 1-2%.

**Laboratory determination of the parameters $C_\lambda$ and $\mu_\lambda$.**

In order to obtain the spectra of absorption corresponding to different physical conditions, we used the VKM-100 multipass cell with the length 96.7 m. The cell is equipped with a system of mirrors according to White's schema, which provides multiple passages of light in the cell, with the minimum step equal to four passages [10]. The detection of the spectra was carried out with the SF-68 spectrophotometer [11], a common instrument for spectrophotometrical observations of stars. The spectra were recorded to punched tape, and, for visual control, with a plotter. As a detector, the uncooled photomultiplier FEU-83 was used. The observations were carried out with the widths of the exit slit of the spectrophotometer 25, 50, 100, and 150 Å. The source of the continuum spectrum was a G-33 incandescent filament lamp with electronic voltage stabilization. The pressure in the cell was measured with standard vacuummeters. A particular attention was given to the determination of the water vapor content in the cell. To this end, the portion of the water vapor spectrum was additionally detected with high spectral resolution, with the ASP-12 spectrograph [12]. Figure 2 presents the schematic diagram of the experimental set.

The experiment was carried out in accordance with the following procedure. Initially, the incandescent lamp was observed through the cell with the evacuated air, for different numbers of passages of light. In this case, the light intensity distribution in the spectrum was specified by the radiation of the lamp, the optical parameters of the cell and spectrophotometer, the number of reflections of light in the cell, and the spectral sensitivity of the detector. Next, the spectra of the same lamp were observed through the cell filled with moisturized air, for different pressure values, starting from atmospheric (P = 1.0, 0.7, 0.5, 0.3, 0.2, and 0.1 atm). The humidity of the air in the cell was initially established with the atmospheric pressure by introduction a certain amount of liquid water to the cell. Later on, for each value of the pressure, the amount of water vapor along the line of sight varied by the variation of the number of passages of light in the cell. We used the following numbers of passages: 5, 9, 13, 17, 21, 25, 29, and 33. A larger number could not be used since the light was substantially attenuated due to multiple reflections. As a rule, on each particular day, observations for some specific pressure were carried out, which had been established on the previous day, and during the subsequent night the humidity distribution along the cell had been homogenized. For each value of pressure, after the registration of the spectra by SF-68 spectrophotometer, observations with ASP-12 were carried out, also for various numbers of passages. The set of measurements finished with another series of observations for the zero pressure (in the evacuated cell).

The water vapor content in the cell was determined using the $\lambda$ 6943.8 Å line of water vapor. From photoelectric records obtained for various numbers of passages, the equivalent width of the 6943.8 Å line was measured. Then, using a theoretical growth curve, the amount of water vapor along the line of sight was derived from the measured equivalent widths. The theoretical growth curve for each pressure was constructed under the assumption that the line contour is specified by collisions and Doppler broadening. The line intensity $S = 0.18$ cm$^{-1}$/g [13] and the half-width $\gamma = 0.09$ cm$^{-1}$/atm [14] were taken as the line parameters. The theoretical equivalent widths were calculated using Tables [15]. The water vapor content in the cell was derived from the experimental dependence of the amount of water vapor on the number of passages (Fig.3). This method makes it possible to determine the water vapor content in the air with the accuracy of 2-3%. The residual amount of water vapor, which in Fig.3 is extrapolated to the zero number of passages, corresponds to the amount of vapor in the spectrograph and in the air layer between the spectrograph and the cell.

The spectra were detected with the SF-68 spectrophotometer in the same mode as that applied in observations of stellar spectra. Therefore for the primary processing (filtering out noise, wavelength calibration, data averaging) we used codes previously composed for the processing of stellar spectra. As a result, we obtained the sets of observed spectra in stellar magnitudes $m_\lambda^{P,\omega}$, which corresponded to different conditions in the cell (in terms of P and $\omega$), and the sets of spectra for the evacuated cell $m_\lambda^{0,i}$ for different numbers of passages (i = 5, 9,..., 33). In order to compensate variations of the transmission in the cell, related to adjustment inaccuracies, the spectra $m_\lambda^{P,\omega}$ and $m_\lambda^{0,i}$ were brought into coincidence within absorption-free wavelength intervals. After this procedure, the value

$$\Delta m_\lambda^{P,\omega} = m_\lambda^{P,\omega} - m_\lambda^{0,i} \qquad (6)$$

is specified only by the absorption of radiation in individual lines of the band. Next, the experimental values of $m_\lambda^{P,\omega}$ were used for determination of the parameters $C_\lambda$ and $\mu_\lambda$ with the use of expression (5), written in logarithmical form for the sake of convenience:

$$\lg \Delta m_\lambda = \lg C_\lambda + \mu_\lambda \cdot \lg \omega . \qquad (7)$$

In the determination of the parameters for given conditions in the cell, we used only the spectra obtained by a variation of the number of passages, for a given pressure and water vapor content. All sets of the parameters displayed comparable accuracy and the same conditions relative to the variation of the amount of the absorbing matter along the line of sight, since the maximum amount of the water vapor along the line of sight always exceeded the minimum by the factor of 6.6 (the ratio of 33 and 5 passages). To test the reliability of the determination of the water vapor content in the cell, and also the accuracy of the position of the continuum spectrum within the absorption band, after bringing the spectra to the coincidence, we measured the total absorption in the bands for each spectrum. It does not depend on the spectral resolution, which makes it possible to compare all data obtained with different widths of the exit slit of the spectrophotometer for a given pressure. The analysis of the data scatter in the graph of the total absorption as a function of the amount of water vapor along the line of sight indicates that the error of determination for the continuum level within the absorption bands is confined within the interval 1-2%, while the error of determination for the water vapor content in the cell does not exceed 3-4%.

**The results of the determination of the empirical parameters.**
As a result of the processing of the experimental data, we obtained approximately 50 sets of the parameters $C_\lambda$ and $\mu_\lambda$ corresponding to different conditions in the cell and values of spectral resolution. We used these data to analyze the adequacy of the approximation of the observations by the formula (5) with the use of empirical parameters $C_\lambda$ and $\mu_\lambda$. If we compare the expression (5) with the description of absorption based on band models [16], then in the case of absorption by weak "unsaturated" lines (an optically thin layer), $\mu_\lambda = 1$, while in the case of absorption by "saturated" lines (an optically thick layer), $\mu_\lambda = 0.5$. Consequently, a set of parameters obtained for a large amount of absorbing matter, when lines are saturated, will not correspond to that obtained with weak absorption, for unsaturated lines. Thereby, the same set of parameters cannot describe experimental data adequately in a very broad interval of variations of the amount of absorbing matter.

For water vapor contents in the atmosphere, typical for astronomical observations, in the center of the 930 nm band most of the lines are saturated, in the wings of the band the number of saturated lines decreaseas, and in 720 and 820 nm bands only the strongest lines indicate some signs of transition to the saturation stage. Therefore, in the center of the 930 nm band, the parameter $\mu_\lambda$ will be close to 0.5, while in the wings of the 930 nm band and in the 720 and 820 nm bands, it will be in the interval between 0.5 and 1.

In addition to that, the parameter $\mu_\lambda$ varies from 1 to 0.5 within the interval of water vapor contents at which the transition from the stage of unsaturated lines to the stage of their saturation in some portion of a band occurs; consequently, within this interval a dependence of parameter $\mu_\lambda$ on the amount of absorbing matter should be expected.

Indeed, Fig.4a, which presents the parameter $\mu_\lambda$ obtained for intervals which differ roughly by the factor of two in terms of the amount of absorbing matter, indicates that the parameter $\mu_\lambda$ varies in the wings of the 930 nm band. It may seem that these speculations, as well as the results presented in Fig. 4a, provide evidence against the adequacy of the expression (5) as a function that approximates the observed absorption variations. However, this is not the case. Taking into account that the interval of water vapor contents in which astronomical observations may be carried out is restricted, precisely the set of parameters obtained for this interval may be recommended as the optimum. Moreover, application of these parameters beyond the region in which they are determined will not result in an immediate loss in accuracy, either; the error will grow only gradually. For example, Fig.5 presents the comparison of absorption values calculated for two sets of parameters, which are best fit for values of the water vapor contents that differ by the factor of about two. It follows from Fig.5 that a noticeable difference in the values of the parameter $\mu_\lambda$ for $\lambda$ 9700 Å results in the difference between calculated values of absorption of 1-2 %, which is close to the accuracy of the experimental data.

The expression (4) suggests that the dependence of absorption on the amount of absorbing matter and pressure may be considered separately for these factors, i.e. the parameter $\mu_\lambda$ does not depend on pressure. Figure 4b, which presents the parameter $\mu_\lambda$ obtained for different values of pressure, confirms this suggestion. A variation of the exit slit width in our experiment does not result in any substantial variation in the spectral dependence of the parameter $\mu_\lambda$, either. Figure 4c presents the parameter $\mu_\lambda$ as a function of $\lambda$ for different exit slit widths. Although with the decrease of the slit width some details appear in the structure of the spectral dependence of the parameter $\mu_\lambda$, these variations are small, and the application of the same set of values for the parameter $\mu_\lambda$ for all values of resolution remains valid. As it was noted above, a variation of the resolution alters not only the numerical value of the parameter $C_\lambda$, which corresponds to absorption for 1 cm of precipitated water, but also the position of individual maxima and minima of the function $C_\lambda$ (Fig.6).

As a result of our analysis of the experimental data, we selected the sets of parameters $C_\lambda$ and $\mu_\lambda$ which may be recommended for the reduction of spectrophotometrical observations. Table1 presents the values of the parameter $C_\lambda$ obtained for the pressure 0.7 atm and corresponding to the width of the exit slit of the spectrophotometer 25, 50, 100, and 150 Å. Since the processing included smoothing of the spectra within the window of 35 Å, the effective resolution will be different in all cases. Numerical experiments with different intervals of averaging indicated that the effective slit width may be obtained as a root of the sum of the squares of the slit width and the averaging interval, which in our case yields the effective slit width of 41, 61, 106, and 154 Å. The error of determination of the parameter $C_\lambda$ does not vary within the band and equals 0.003-0.005. Table 1 contains parameter $\mu_\lambda$ averaged over different values of resolution, along with its determination error. In the wings of the bands, where the absorption is small, the parameter $\mu_\lambda$ is determined unreliably; in these cases, it is taken to be a unity. For Table 1, we selected the values of the parameters obtained within the interval of water vapor contents 0.3-3.5 cm of precipitated water, as the most typical in astronomical observations. The pressure 0.7 atm is also selected as the most characteristic for a mountain observatory; it corresponds to the effective pressure of water vapor at the height exceeding 2000 m above the sea level.

Sets of parameters $C_\lambda$ and $\mu_\lambda$ similar to those presented in Table1 were also obtained for the pressure 0.5 and 1.0 atm, and in addition, for the exit slit width of 50 and 100 Å, for the pressure 0.1, 0.2, and 0.3 atm. The sets of the parameters for different pressure values were used for the analysis of the dependence of the parameter $C_\lambda$ on pressure, which, according to the expression (5), may be presented by a power function with an empirically determined power index $n_\lambda$. Within the considered interval of pressure, this function approximates the experimental data quite satisfactorily (Fig.7). Table 2 presents the results of the determination of the parameter $n_\lambda$, along with its determination error, for the 930 nm band. The dependence of the parameter $n_\lambda$ on the width of the slit of the spectrophotometer may be neglected, like it was in the case of the parameter $\mu_\lambda$. For the 720 and 820 nm bands, the differences in the results of the determination of the parameter $n_\lambda$ for different wavelengths are too large to establish the form of the wavelength dependence. For the considered bands, the average value $n_\lambda = 0.44$ be recommended.

Thereby, the empirical parameters presented in Tables 1 and 2 make it possible to form a set of parameters $C_\lambda$ and $\mu_\lambda$ for a specific place of observations (for some effective pressure) and a specific spectral device (for some spectral resolution). Using these parameters, for a known water vapor content along the line of view, the attenuation in water vapor absorption bands may be calculated with the accuracy $0.^m01$-$0.^m02$, which provides the same accuracy for the correction of the observed spectra for the atmospheric absorption.

The authors thank L.A.Kamionko, E.S.Kulagina, A.H.Kurmayeva, I.N.Nikanorova, V.P.Pakhomov, and Yu.N.Chistyakov for useful discussions.

Empirical parameters $C_\lambda$ and $\mu_\lambda$. **Table 1.**

| $\lambda$, Å | $C_\lambda$ | | | | $\mu_\lambda$ | $\Delta\mu_\lambda$ |
| --- | --- | --- | --- | --- | --- | --- |
| | 25 Å | 50 Å | 100 Å | 150 Å | | |
| 6925 | 0.018 | 0.010 | 0.008 | 0.017 | 1.000 | |
| 6950 | 0.024 | 0.019 | 0.008 | 0.018 | 1.000 | |
| 6975 | 0.025 | 0.018 | 0.017 | 0.020 | 1.000 | |
| 7000 | 0.033 | 0.023 | 0.019 | 0.023 | 1.000 | |
| 7025 | 0.025 | 0.024 | 0.014 | 0.017 | 1.000 | |
| 7050 | 0.025 | 0.021 | 0.017 | 0.017 | 1.000 | |
| 7075 | 0.022 | 0.016 | 0.015 | 0.016 | 1.000 | |
| 7100 | 0.015 | 0.015 | 0.014 | 0.015 | 1.000 | |
| 7125 | 0.014 | 0.017 | 0.020 | 0.032 | 0.980 | 0.333 |
| 7150 | 0.034 | 0.035 | 0.040 | 0.043 | 0.900 | 0.069 |
| 7175 | 0.125 | 0.090 | 0.065 | 0.060 | 0.858 | 0.055 |
| 7200 | 0.108 | 0.107 | 0.093 | 0.077 | 0.808 | 0.055 |
| 7225 | 0.091 | 0.092 | 0.102 | 0.094 | 0.824 | 0.047 |
| 7250 | 0.107 | 0.099 | 0.091 | 0.092 | 0.878 | 0.053 |
| 7275 | 0.099 | 0.097 | 0.083 | 0.072 | 0.923 | 0.068 |
| 7300 | 0.079 | 0.078 | 0.067 | 0.059 | 0.954 | 0.103 |
| 7325 | 0.044 | 0.042 | 0.047 | 0.040 | 0.966 | 0.154 |
| 7350 | 0.015 | 0.017 | 0.024 | 0.023 | 0.981 | 0.218 |
| 7375 | 0.007 | 0.010 | 0.010 | 0.005 | 1.000 | |
| 7400 | 0.004 | 0.003 | 0.005 | 0.004 | 1.000 | |
| 8000 | 0.009 | 0.007 | 0.011 | 0.011 | 1.000 | |
| 8025 | 0.015 | 0.010 | 0.011 | 0.013 | 1.000 | |
| 8050 | 0.011 | 0.012 | 0.011 | 0.015 | 1.000 | |
| 8075 | 0.010 | 0.008 | 0.011 | 0.023 | 1.000 | |
| 8100 | 0.011 | 0.013 | 0.018 | 0.035 | 0.987 | 0.202 |
| 8125 | 0.032 | 0.043 | 0.045 | 0.047 | 0.909 | 0.078 |
| 8150 | 0.106 | 0.094 | 0.078 | 0.063 | 0.830 | 0.055 |
| 8175 | 0.126 | 0.120 | 0.102 | 0.086 | 0.765 | 0.059 |
| 8200 | 0.090 | 0.108 | 0.110 | 0.093 | 0.764 | 0.052 |
| 8225 | 0.100 | 0.102 | 0.096 | 0.103 | 0.730 | 0.062 |
| 8250 | 0.083 | 0.089 | 0.086 | 0.090 | 0.755 | 0.064 |
| 8275 | 0.073 | 0.077 | 0.080 | 0.073 | 0.748 | 0.078 |
| 8300 | 0.072 | 0.074 | 0.065 | 0.063 | 0.781 | 0.096 |
| 8325 | 0.055 | 0.055 | 0.051 | 0.042 | 0.809 | 0.166 |
| 8350 | 0.024 | 0.034 | 0.033 | 0.031 | 0.921 | 0.180 |
| 8375 | 0.014 | 0.017 | 0.017 | 0.021 | 0.990 | 0.292 |
| 8400 | 0.007 | 0.010 | 0.008 | 0.009 | 1.000 | |
| 8425 | 0.006 | 0.006 | 0.006 | 0.006 | 1.000 | |
| 8450 | 0.006 | 0.006 | 0.005 | 0.005 | 1.000 | |

**Table 1 (cont.).**

Empirical parameters $C_\lambda$ and $\mu_\lambda$.

| λ, Å | $C_\lambda$ | | | | $\mu_\lambda$ | $\Delta\mu_\lambda$ |
| --- | --- | --- | --- | --- | --- | --- |
| | 25 Å | 50 Å | 100 Å | 150 Å | | |
| 8925 | 0.016 | 0.022 | 0.038 | 0.065 | 0.847 | 0.110 |
| 8950 | 0.065 | 0.070 | 0.095 | 0.082 | 0.792 | 0.066 |
| 8975 | 0.156 | 0.158 | 0.135 | 0.109 | 0.700 | 0.049 |
| 9000 | 0.190 | 0.171 | 0.140 | 0.130 | 0.668 | 0.051 |
| 9025 | 0.136 | 0.132 | 0.142 | 0.153 | 0.701 | 0.037 |
| 9050 | 0.099 | 0.120 | 0.136 | 0.162 | 0.743 | 0.036 |
| 9075 | 0.153 | 0.151 | 0.148 | 0.166 | 0.723 | 0.031 |
| 9100 | 0.177 | 0.182 | 0.175 | 0.172 | 0.733 | 0.026 |
| 9125 | 0.188 | 0.191 | 0.180 | 0.173 | 0.747 | 0.029 |
| 9150 | 0.198 | 0.194 | 0.162 | 0.158 | 0.752 | 0.038 |
| 9175 | 0.161 | 0.146 | 0.135 | 0.137 | 0.792 | 0.050 |
| 9200 | 0.102 | 0.103 | 0.108 | 0.133 | 0.857 | 0.046 |
| 9225 | 0.086 | 0.087 | 0.110 | 0.167 | 0.883 | 0.032 |
| 9250 | 0.101 | 0.117 | 0.186 | 0.253 | 0.790 | 0.027 |
| 9275 | 0.209 | 0.245 | 0.330 | 0.361 | 0.632 | 0.024 |
| 9300 | 0.453 | 0.490 | 0.526 | 0.470 | 0.573 | 0.018 |
| 9325 | 0.765 | 0.755 | 0.707 | 0.565 | 0.556 | 0.010 |
| 9350 | 0.943 | 0.920 | 0.764 | 0.648 | 0.561 | 0.010 |
| 9375 | 0.785 | 0.770 | 0.723 | 0.705 | 0.581 | 0.011 |
| 9400 | 0.562 | 0.625 | 0.655 | 0.690 | 0.608 | 0.009 |
| 9425 | 0.588 | 0.605 | 0.613 | 0.623 | 0.607 | 0.010 |
| 9450 | 0.666 | 0.678 | 0.633 | 0.598 | 0.594 | 0.010 |
| 9475 | 0.643 | 0.645 | 0.632 | 0.598 | 0.595 | 0.010 |
| 9500 | 0.602 | 0.619 | 0.601 | 0.582 | 0.601 | 0.011 |
| 9525 | 0.567 | 0.589 | 0.562 | 0.541 | 0.614 | 0.011 |
| 9550 | 0.536 | 0.553 | 0.522 | 0.499 | 0.619 | 0.011 |
| 9575 | 0.596 | 0.512 | 0.475 | 0.452 | 0.626 | 0.012 |
| 9600 | 0.439 | 0.453 | 0.419 | 0.378 | 0.649 | 0.016 |
| 9625 | 0.375 | 0.367 | 0.332 | 0.306 | 0.660 | 0.023 |
| 9650 | 0.299 | 0.290 | 0.246 | 0.249 | 0.675 | 0.026 |
| 9675 | 0.181 | 0.188 | 0.189 | 0.201 | 0.716 | 0.024 |
| 9700 | 0.110 | 0.135 | 0.150 | 0.161 | 0.764 | 0.028 |
| 9725 | 0.129 | 0.138 | 0.135 | 0.129 | 0.766 | 0.025 |
| 9750 | 0.154 | 0.156 | 0.135 | 0.110 | 0.746 | 0.030 |
| 9775 | 0.128 | 0.130 | 0.116 | 0.097 | 0.763 | 0.011 |
| 9800 | 0.092 | 0.095 | 0.088 | 0.078 | 0.801 | 0.062 |
| 9825 | 0.059 | 0.058 | 0.055 | 0.051 | 0.885 | 0.093 |
| 9850 | 0.028 | 0.034 | 0.035 | 0.027 | 0.950 | 0.253 |
| 9875 | 0.010 | 0.018 | 0.019 | 0.012 | 1.000 | |
| 9900 | 0.003 | 0.009 | 0.009 | 0.007 | 1.000 | |

Empirical parameter $n_\lambda$.    **Table 2.**

| $\lambda$, Å | $n_\lambda$ | $\Delta n_\lambda$ | $\lambda$, Å | $n_\lambda$ | $\Delta n_\lambda$ |
|---|---|---|---|---|---|
| 8950 | 0.47 | 0.05 | 9400 | 0.46 | 0.01 |
| 8975 | 0.51 | 0.02 | 9425 | 0.46 | 0.01 |
| 9000 | 0.56 | 0.03 | 9450 | 0.45 | 0.01 |
| 9025 | 0.52 | 0.03 | 9475 | 0.45 | 0.01 |
| 9050 | 0.50 | 0.04 | 9500 | 0.44 | 0.01 |
| 9075 | 0.46 | 0.02 | 9525 | 0.44 | 0.01 |
| 9100 | 0.42 | 0.03 | 9550 | 0.44 | 0.01 |
| 9125 | 0.41 | 0.03 | 9575 | 0.43 | 0.01 |
| 9150 | 0.41 | 0.02 | 9600 | 0.42 | 0.01 |
| 9175 | 0.40 | 0.02 | 9625 | 0.30 | 0.01 |
| 9200 | 0.40 | 0.05 | 9650 | 0.41 | 0.02 |
| 9225 | 0.34 | 0.05 | 9675 | 0.39 | 0.01 |
| 9250 | 0.35 | 0.05 | 9700 | 0.51 | 0.02 |
| 9275 | 0.38 | 0.05 | 9725 | 0.43 | 0.02 |
| 9300 | 0.41 | 0.01 | 9750 | 0.44 | 0.02 |
| 9325 | 0.39 | 0.05 | 9775 | 0.49 | 0.02 |
| 9350 | 0.44 | 0.01 | 9800 | 0.48 | 0.05 |
| 9375 | 0.45 | 0.01 | 9825 | 0.47 | 0.02 |

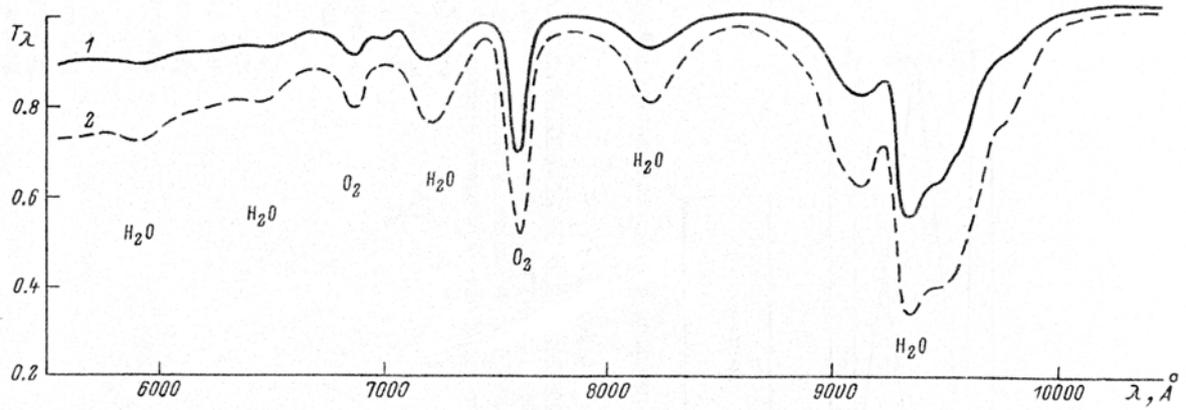

**Fig. 1**. Atmospheric transmission for height 2000 m and water vapor content 2 cmppw.
**1** – air mass = 1;    **2**- air mass = 3

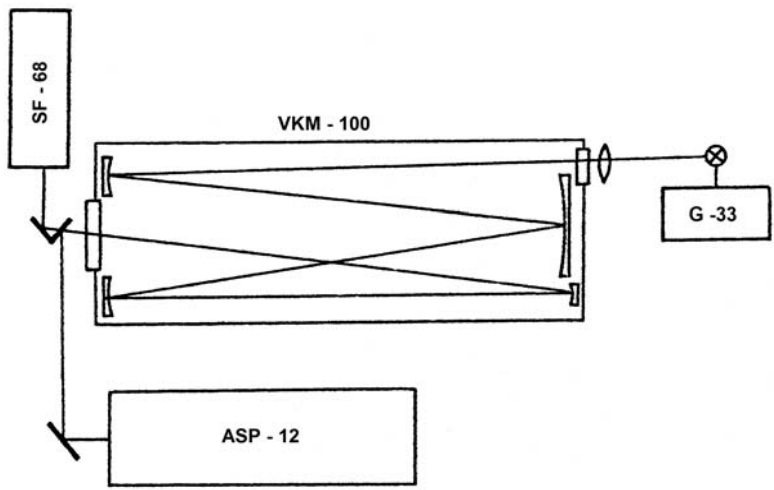

**Fig. 2**. The schematic diagram of the experimental set.

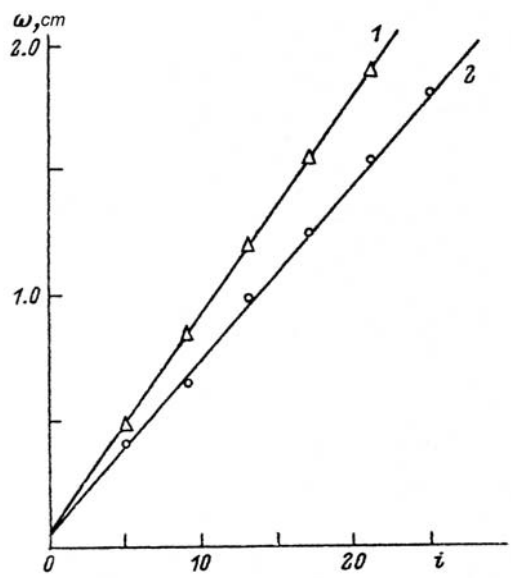

**Fig. 3**. The determination of water vapor content in the cell.
**1** – 28.12.1984, P = 1 atm;   **2** - 06.01.1985, P = 0.2 atm;
i – number of passages.

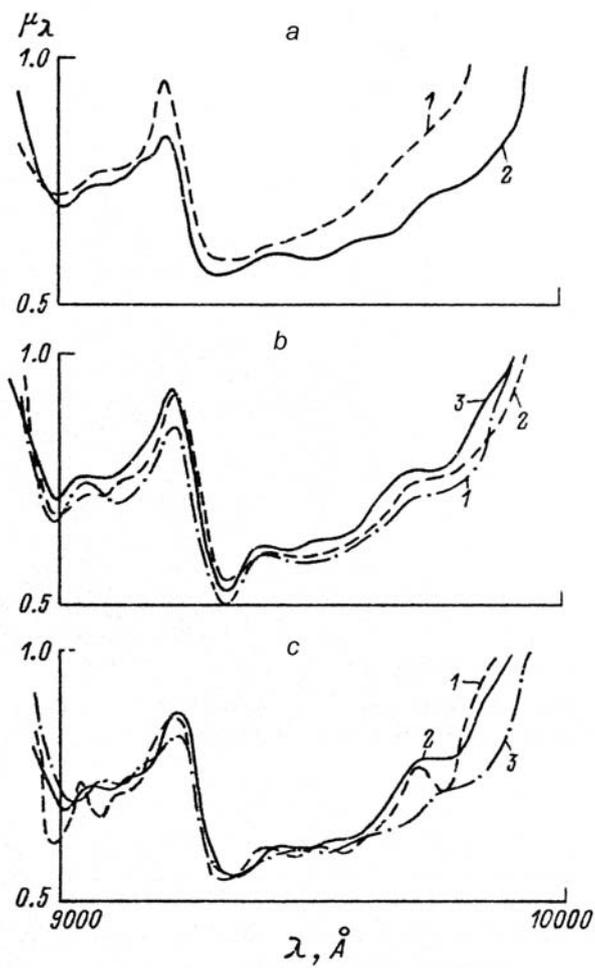

**Fig. 4**. Empirical parameter $\mu_\lambda$ for water vapor band 0.93 mcm.
**a** – P = 0.7 atm; **1** – water vapor content along the line of view 0.24-1.6, **2** – 0.44-2.87 cmppw;
**b** – exit slit width 50 Å; **1** – P = 0.5, **2** – P = 0.7, **3** – P = 1.0 atm.
**c** – P = 0.7 atm; exit slit width: **1** – 25 Å, **2**- 50 Å, **3** – 100 Å.

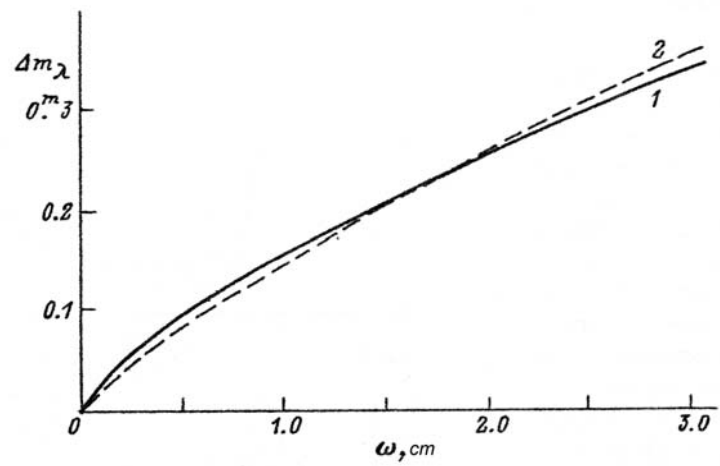

**Fig. 5**. Absorption dependence at λ 9700 Å from water vapor content along the line of view.
**1** – $C_\lambda = 0.^m157$, $\mu_\lambda = 0.706$; **2** - $C_\lambda = 0.^m145$, $\mu_\lambda = 0.834$

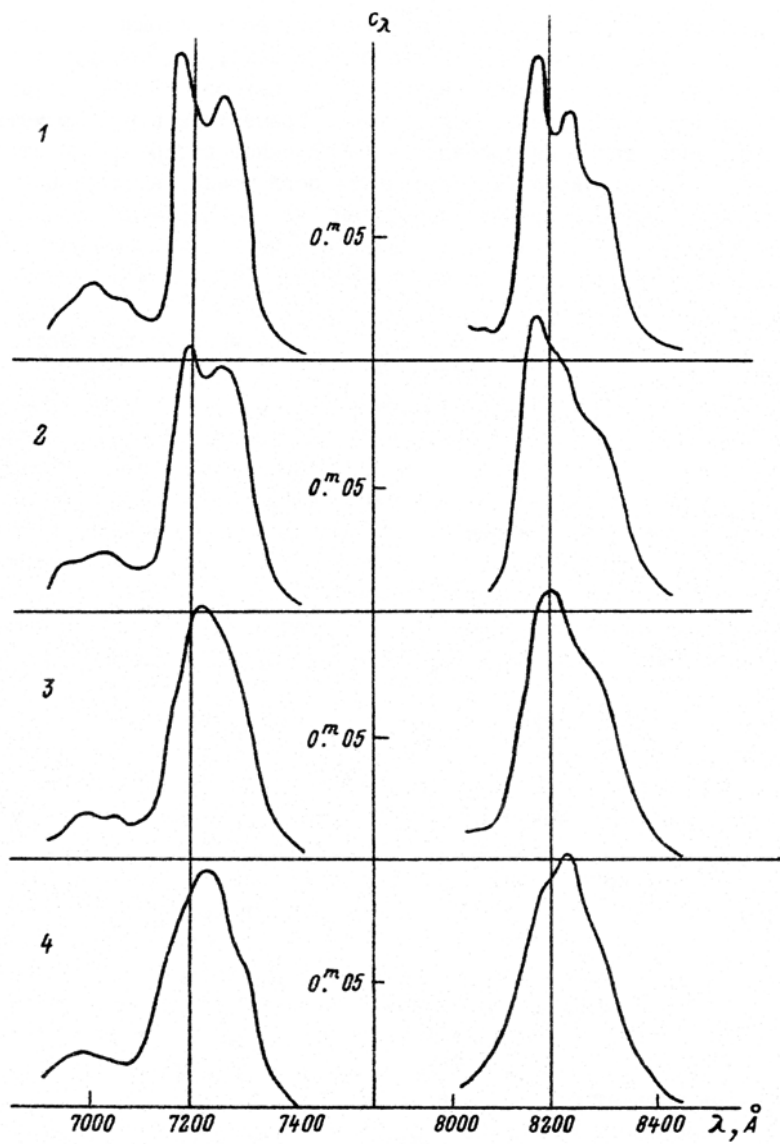

**Fig. 6**. Empirical parameter $C_\lambda$ for water vapor bands 0.72 mcm (at the left) and 0.82 mcm (on the right). Exit slit width: **1** – 25 Å, **2** – 50 Å, **3** - 100 Å, **4** – 150 Å.

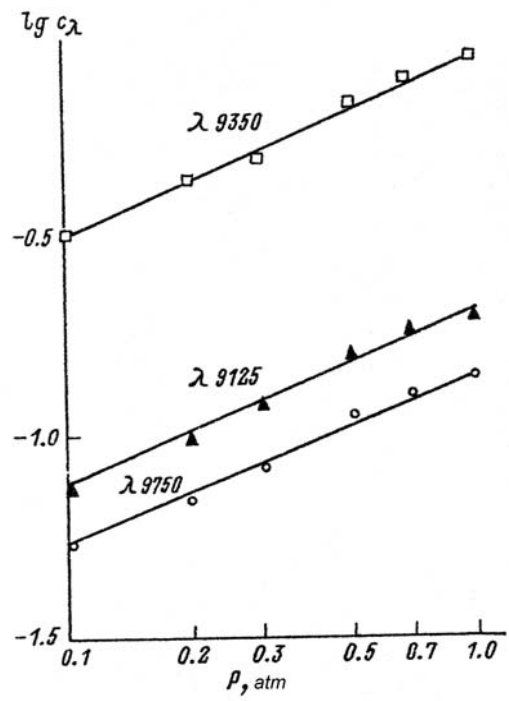

**Fig. 7**. The determination of empirical parameter $n_\lambda$.